\documentclass[aps,showpacs,superscriptaddress,floatfix,reprint]{revtex4-1}
\usepackage{bm}
\usepackage{amsmath}
\usepackage{amssymb}
\usepackage{amsthm}
\usepackage{graphicx}
\usepackage{cases}

\usepackage{graphicx}
\usepackage{epstopdf}

\begin{document}

\title{Multiple-scale approach for the expansion scaling of superfluid quantum gases}

\author{I. L. Egusquiza}
\affiliation{Department of Theoretical Physics and History of Science, UPV-EHU, 48080 Bilbao, Spain}
\author{M. Modugno}
\affiliation{Department of Theoretical Physics and History of Science, UPV-EHU, 48080 Bilbao, Spain}
\affiliation{IKERBASQUE, Basque Foundation for Science, 48011 Bilbao, Spain}
\author{ M. A. Valle Basagoiti}
\affiliation{Department of Theoretical Physics and History of Science, UPV-EHU, 48080 Bilbao, Spain}

\date{\today}

\begin{abstract}
We present a general method, based on a \textit{multiple-scale} approach, for deriving the   perturbative solutions of the scaling equations governing the expansion of superfluid ultracold quantum gases released from elongated harmonic traps. We discuss how to treat the secular terms appearing in the usual naive expansion in the trap asymmetry parameter $\epsilon$, and calculate the next-to-leading correction for the asymptotic aspect ratio, with significant improvement over the previous proposals.
 
\end{abstract}

\maketitle

\section{Introduction}

The physics of ultracold quantum gases has been the object of an intense experimental and theoretical investigation since the achievement of Bose-Einstein condensation in dilute trapped gases in 1995 \cite{dalfovo,giorgini}. A common ingredient of these experiments is the so called time-of-flight, that is the expansion of the gas after the release of the confining potential.
For some typical experimental regimes, the expansion after the release from harmonic traps can be described by scaling solutions, in terms of three scaling parameters $b_i(t)$ ($i=1,2,3$), which obey a system of second-order ordinary differential equations, of the general form (where $\omega_{i}$ are the trapping frequencies)
\cite{castin,kagan,dalfovo,menotti,giorgini,schafer}
\begin{equation}
\ddot{b_i}=\frac{\omega_i^2}{{(b_x b_y b_z)}^\gamma}\frac{1}{b_i}
\end{equation}
with initial conditions $b_{i}(0)=1$, $\dot{b}_{i}(0)=0$ . In particular, $\gamma=1$ corresponds to the case of a Bose-Einstein condensate (BEC) in the Thomas-Fermi (TF) limit \cite{castin,kagan,dalfovo}, whereas $\gamma=2/3$ refers to a unitary superfluid Fermi gas \cite{menotti,giorgini,schafer}.
The solution of the above equations allows the complete characterization of the expansion of the cloud size as $R_i(t)=R_i(0)b_i(t)$.

In the case of strongly elongated cylindrically-symmetric traps, $\epsilon=\omega_z/\omega_\perp\ll 1$, the above equations read 
\begin{numcases}{}
\label{eq:bz}
\frac{d^2 b_{\perp}}{d\tau^2} = \frac{1}{ b_{\perp}^{2\gamma+1}  b_z^{\gamma}} \\
\frac{d^2 b_z}{d\tau^2} = \frac{\epsilon^2}{b_{\perp}^{2\gamma}b_z^{\gamma+1}}.
\label{eq:bperp}
\end{numcases}
with $\tau\equiv\omega_\perp t$.
Usually, these equations are solved by means of an expansion in powers of  $\epsilon^{2}$, by the ansatz 
\begin{equation}
b_i(\tau) = \sum_{n=0}^\infty \epsilon^{2n}  b_{i}^{(n)}(\tau).
\label{eq:naive}
\end{equation} 
For example, for the BEC case $\gamma=1$, one retains the zeroth-order in $b_\perp$, and up to second order in $b_z$, leading to the well-known Castin and Dum scaling \cite{castin}
\begin{numcases}{}
 b_{\perp}(\tau) = \sqrt{1+\tau^2} \\
 b_z(\tau) = 1 + \epsilon^2\left(\tau\arctan\tau - \ln\sqrt{1+\tau^2}\right)
 \label{eq:bperp_cd}
\end{numcases}
that nicely describes the experimental data \cite{dalfovo}.

From these expression one can also infer the asymptotic aspect ratio, defined as $\lim_{\tau\to\infty}(R_\perp(\tau)/R_z(\tau))=(R_\perp(0)/R_z(0))(2/\pi\epsilon^2)$ \cite{dalfovo}.
However, one may notice that the above expansion is strictly valid only in the perturbative regime, when the second order correction in (\ref{eq:bperp_cd}) is smaller that unity, namely for $\tau\ll2/\pi\epsilon^2$. This condition is in general well satisfied in the typical experimental regimes, but in principle does not permit the  direct extraction of  the asymptotic limit, because of the secular term $\tau\arctan\tau$ that eventually invalidates the perturbative expansion. 

Here we present a general method, based on a \textit{multiple-scale} perturbative expansion, that allows the derivation of a \textit{uniformly valid} expansion, where the hierarchy of sub-leading terms is preserved at \textit{any time}. This can be achieved by means of a proper resummation of the secular terms \cite{multiplescale}.

The paper is organized as follows. In Sect. \ref{sect:formulation} we reformulate the problem by using a Hamiltonian approach (see also \cite{kagan}). In \S\ref{sect:naiveA} we show that the leading term of the asymptotic ratio can be computed already at zeroth-order in the naive expansion, by using the expression in terms of the canonical momenta (avoiding secularities). We find that this result may represent a good approximation of the exact (numerical) asymptotic aspect ratio, depending on the value of $\gamma$, and provided that $\epsilon$ is small enough. However, it  may deviate significantly from the exact solution for $\gamma\lesssim1$, already for not too large values of $\epsilon$ (e.g. $\epsilon\approx0.05$). In \S\ref{sect:naiveB} we show that the next-to-leading order in the naive expansion gives rise to secular terms for any $\gamma$, requiring therefore a different expansion approach.
Then, in Sect. \ref{sect:multiscale} we introduce the general ideas for the multiple scale approach, and consider the next-to-leading correction to the asymptotic aspect ratio, addressing in particular the case $\gamma=1$ and $1/2<\gamma<1$. In both cases the expansion tuns out to be \textit{non-analytic} due to the presence of terms proportional to $\epsilon^{2}\log\epsilon$  ($\gamma=1$) or  ${\epsilon}^{2(2 \gamma-1)}$ ($\gamma<1$).
We then recapitulate and offer prospective applications of the method.

\section{Formulation of the problem}
\label{sect:formulation}
\begin{figure*}[t!]
\includegraphics[width=0.6\columnwidth]{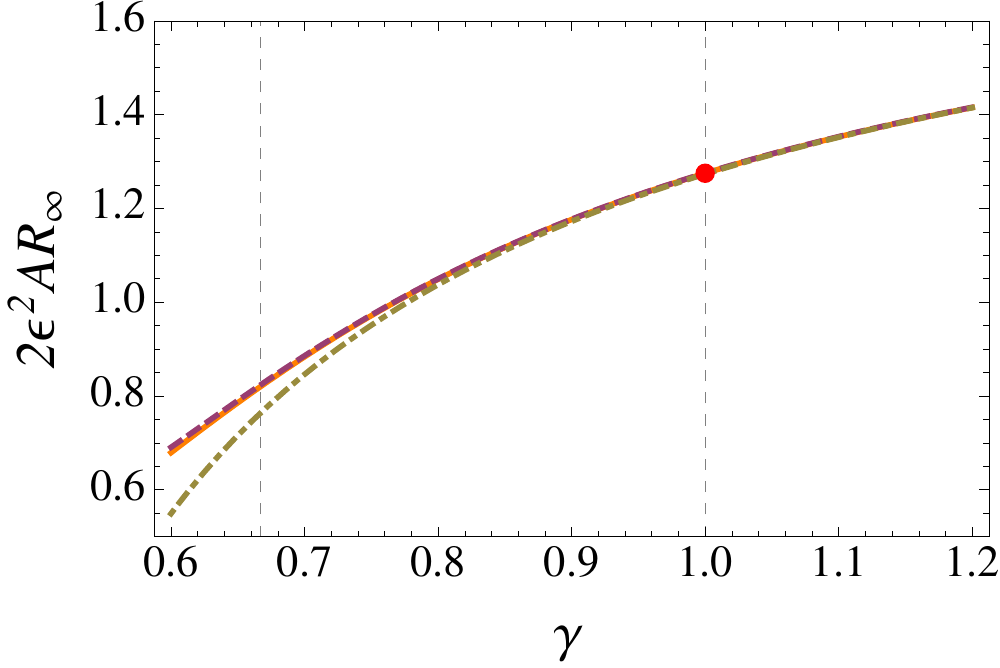}
\includegraphics[width=0.6\columnwidth]{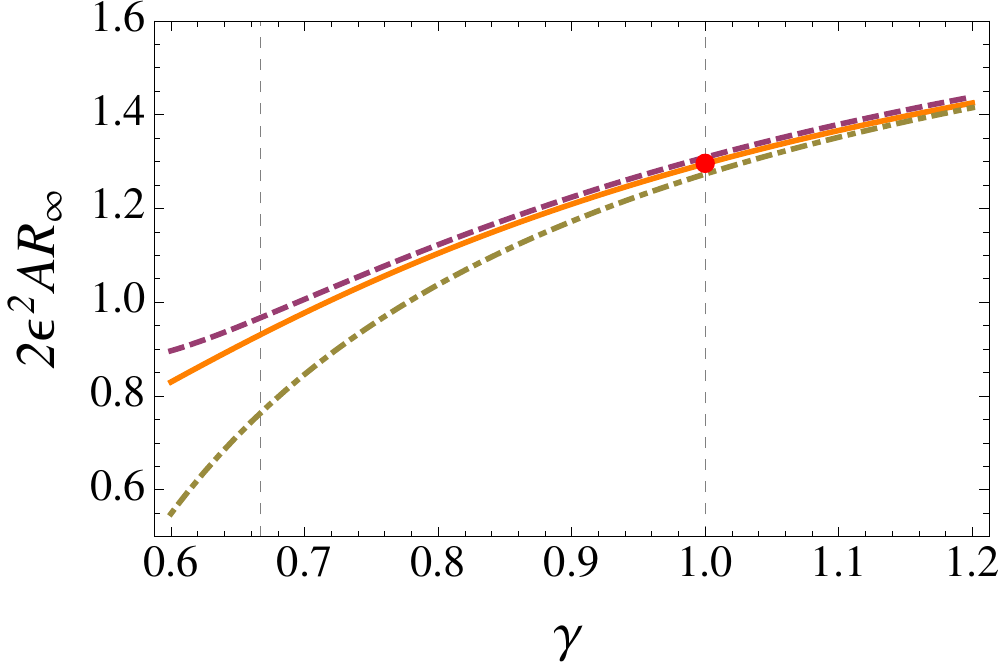}
\includegraphics[width=0.6\columnwidth]{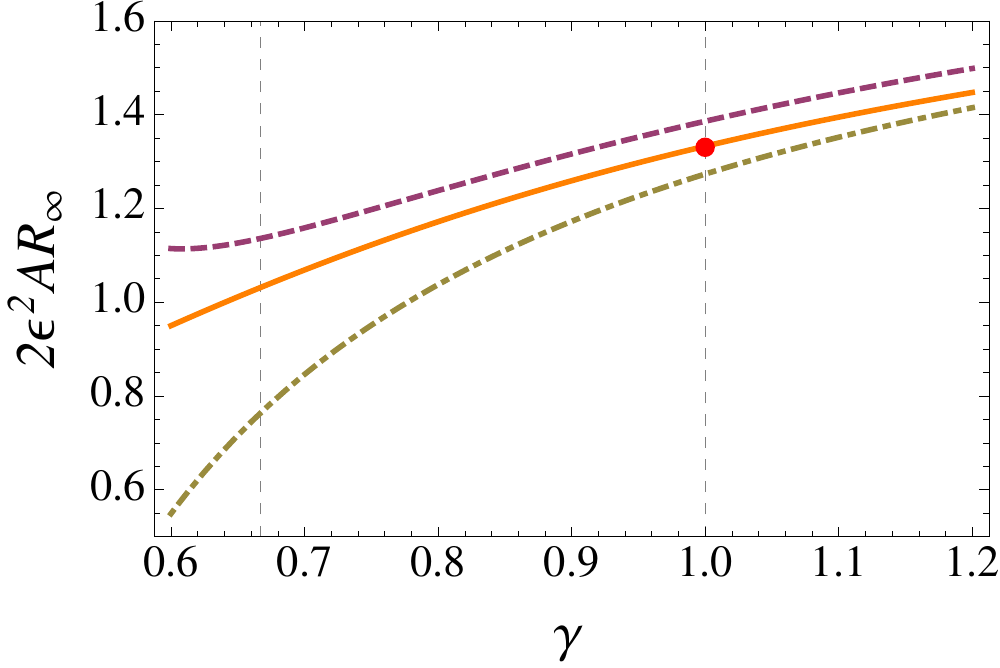}
\caption{(Color online) Asymptotic aspect ratio $2\epsilon^2 AR_{\infty}$ as a function of $\gamma$, for $\epsilon=0.01,0.05,0.1$ (from left to right). The full numerical solution of Eqs. (\ref{eq:bz})-(\ref{eq:bperp}) (orange solid line) is compared with the zeroth-order expression in Eq. (\ref{eq:asympt}) (green dot-dashed line) and with that including the next-to-leading corrections (magenta dashed line, red dot at $\gamma=1$) discussed in Sect. \ref{sect:multiscale} .
The vertical (dashed) lines correspond to $\gamma=2/3$ and $\gamma=1$.}
\label{fig:asympt}
\end{figure*}

Equations (\ref{eq:bz})-(\ref{eq:bperp}) are equivalent to the canonical equations 
($\dot{b}_i={\partial H}/{\partial p_i}$, $\dot{p}_i=-{\partial H}/{\partial b_i}$)
\begin{eqnarray}
&&\dot{b}_\perp=p_\perp
\qquad
\dot{p}_\perp=\frac{1}{b_{\perp}^{2\gamma+1}b_z^{\gamma}}
\\
\label{eq:bzdot}
&&\dot{b}_z=2\epsilon^2p_z
\qquad
\dot{p}_z=\frac{1}{2 b_{\perp}^{2\gamma}b_z^{\gamma+1}}\,.
\end{eqnarray}
generated by the following Hamiltonian \cite{kagan}
\begin{equation}
H=\frac{1}{2}p_\perp^2+{\epsilon^2}p_z^2+\frac{1}{2\gamma b_\perp^{2\gamma} b_z^{\gamma}},
\end{equation}
where the initial conditions $b_{i}(0)=1$, $\dot{b}_{i}(0)=0$ set the total energy to $H=1/2\gamma$. Asymptotically, $\ddot{b}_i\approx0$ and therefore $b_i\propto \tau$, so that $AR_{\infty}\equiv\lim_{\tau\to\infty}(b_\perp/b_z)=(1/2\epsilon^2)(p_\perp/p_z)_{\infty}$.
 
\subsection{Zeroth-order of the naive expansion}
\label{sect:naiveA}

Let us start by considering the naive expansion in Eq. (\ref{eq:naive}), both for $b_{i}(\tau)$ and $p_{i}(\tau)$. 
To lowest order in $\epsilon^2$, the Hamilton equations are
\begin{eqnarray}
&&\dot{b}^{(0)}_{\perp}=p^{(0)}_\perp
\qquad
\dot{p}^{(0)}_\perp=\frac{1}{b^{(0)^{2\gamma+1}}_{\perp}b^{(0)^{\gamma}}_z}
\\
&&\dot{b}^{(0)}_z=0
\qquad
\dot{p}^{(0)}_z=\frac{1}{2 b^{(0)^{2\gamma}}_{\perp}b^{(0)^{\gamma+1}}_z}\,.
\end{eqnarray}
Then, the zeroth-order solution is simply $b^{(0)}_{z}=1$, and the equation for $b_{\perp}^{0}$ becomes 
\begin{equation}
\ddot{b}^{(0)}_{\perp}=\frac{1}{b_{\perp}^{(0)^{2\gamma+1}}}. 
\end{equation}
From energy conservation we have
\begin{equation}
p^{(0)}_\perp=\frac{1}{\sqrt{\gamma}}\sqrt{1-{{b^{(0)}_{\perp}}^{-2\gamma}}}\, ,
\label{eq:p0perp}
\end{equation}
and $b^{(0)}_{\perp}$\  may be easily obtained  in implicit form as
\begin{equation}
b^{(0)}_{\perp}\, _2F_1\left[\frac{1}{2},-\frac{1}{2\gamma};1-\frac{1}{2\gamma};\frac{1}{{b^{(0)}_{\perp}}^{2\gamma}}\right]=\frac{\tau}{\sqrt{\gamma}}+
\frac{(\gamma-1)c(\gamma)}{\sqrt{\gamma}} \, . 
\label{eq:b0perp}
\end{equation}
Here we  have defined  for latter use 
\begin{equation*}
c(\gamma)=\frac{\sqrt{\pi\gamma}}{(\gamma-1)}\frac{\Gamma\left(\frac{2\gamma-1}{2\gamma}\right)}{\Gamma\left(\frac{\gamma-1}{2\gamma}\right)}\,.
\end{equation*}
The asymptotic value of $p^{(0)}_z$ is given by the integral 
\begin{eqnarray}
\lim_{\tau\to\infty}p^{(0)}_{z}  &=&\int_{0}^{\infty}d\tau\dot{p}^{(0)}_{z}(\tau)
\nonumber\\
&=&\int_0^\infty \frac{\sqrt{\gamma}\, dx}{2 x^{2\gamma} \sqrt{1-x^{-2\gamma}}}  =  
\frac{c(\gamma)}{2}
 \label{eq:p0zdot}
\end{eqnarray}
with $\gamma>1/2$; here we have used the fact that $\dot{p}^{(0)}_z={b^{(0)}_{\perp}}\dot{p}^{(0)}_\perp/2$, combined with Eq. (\ref{eq:p0perp}).  From the same equation we get
\begin{equation}
\label{eq:pz0}
\lim_{\tau\to\infty}p^{(0)}_{\perp} = 1/\sqrt{\gamma}\,.
\end{equation}
Therefore, within this hamiltonian formalism, the leading term of the asymptotic ratio can be computed already at zeroth-order by using the expression in terms of the canonical momenta, $AR_{\infty}=(1/2\epsilon^2)(p_\perp/p_z)_{\infty}$ \cite{nota}. 
This has to be compared with the usual approach in terms of the $b_{i}$, where one should include terms up to  order $\epsilon^{2}$ in $b_{z}$ (see Eq. (\ref{eq:bzdot}) and \cite{castin}). By using the above leading values of the momenta we obtain 
\begin{equation}
AR_{\infty}=\frac{1}{\epsilon^{2}}\frac{1}{\sqrt{\gamma}c(\gamma)}=\frac{1}{\epsilon^{2}}\frac{(\gamma-1)}{\sqrt{\pi}\gamma}\frac{\Gamma(1/2-1/2\gamma)}{\Gamma(1-1/2\gamma)}.
\label{eq:asympt}
\end{equation}
From this expression we can extract the TF bosonic case $\gamma=1$, and the unitary fermionic case, $\gamma=2/3$
\begin{numcases}{AR_{\infty}=}
\frac{2}{{\pi}\epsilon^{2}}&$\gamma=1$ \label{eq:gamma1-0}\\
\frac{1}{2\sqrt{\pi}\epsilon^{2}}\frac{-\Gamma[-1/4]}{\Gamma[1/4]}&$\gamma={2}/{3}$
\end{numcases}
that coincide with the results in \cite{dalfovo,shlyap}.

In Fig. \ref{fig:asympt} we compare this zeroth-order prediction (dot-dashed line) with the asymptotic aspect ratio obtained from the full numerical solution \cite{numerics} of Eqs. (\ref{eq:bz})-(\ref{eq:bperp}) (solid line), as a function of $\gamma$. This figure shows that the zeroth-order result (\ref{eq:asympt}) is rather good for $\gamma\gtrsim1$, but deviates significantly from the exact (numerical) solution for lower values of $\gamma$, when $\epsilon$ is not sufficiently small.  We will see in the following how to improve this result.

Notice also that we have not yet implemented the cancellation of secularities (that in fact occurs at the next order in the expansion).

\subsection{Next-to-leading order (naive)}
\label{sect:naiveB}

The next-to-leading order equations are
\begin{eqnarray}
\dot{b}^{(1)}_{\perp}&=&p^{(1)}_\perp
\\
\dot{p}^{(1)}_\perp&=&
-\gamma\frac{b^{(1)}_{z}}{{b^{(0)}}^{2\gamma+1}_{\perp}}
-(2\gamma+1)\frac{b^{(1)}_{\perp}}{{b^{(0)}_{\perp}}^{2\gamma+2}}
\\
\dot{b}^{(1)}_z&=&2 p^{(0)}_z
\label{eq:b1zdot}
\\
\dot{p}^{(1)}_z&=&
-\frac{(1+\gamma)}{2}\frac{b^{(1)}_{z}}{{b^{(0)}_{\perp}}^{2\gamma}}
-\gamma \frac{b^{(1)}_{\perp}}{{b^{(0)}_{\perp}}^{2\gamma+1}}
\end{eqnarray}
Let us consider Eq. (\ref{eq:b1zdot}); by using Eq.~(\ref{eq:p0zdot}) 
the asymptotics is  $\dot{b}^{(1)}_z \sim c(\gamma)$,  
and the corresponding large time  behavior of the axial scaling parameter 
\begin{equation}
b_{z}(\tau)\sim 1+ \epsilon^{2}c(\gamma)\tau, \qquad \tau\to\infty
\end{equation}
grows unboundedly  with $\tau$. This behavior is  physically correct because we expect non-zero asymptotic limits for  both velocities.  
However, with regard to perturbation theory, 
the growing of the first order correction 
must be interpreted as the occurrence of a secular term relative to the naive zeroth-order solution $b_{z}^{(0)}=1$. That is, the perturbative hypothesis breaks down after times of order $1/\epsilon^2c(\gamma)$.
Therefore, to avoid the appearance of these terms 
we propose  a reinterpretation of the perturbative analysis in the 
framework of multiple-scale analysis. 
It is notable that the resummation  involved in this procedure  implies that the higher-order terms of the velocities 
are $O\left((\epsilon^2)^{2\gamma-1}\right)$ 
if $1/2 <\gamma < 1$, and $O(\epsilon^4 \ln \epsilon)$ when $\gamma = 1$. 

\section{Multiple scale approach}
\label{sect:multiscale}

We begin by  introducing an additional time scale $\tau_{1}(\tau)=\epsilon^{2}\tau$, 
and assume a perturbative expansion
\begin{eqnarray}
b_i(\tau,\tau_{1})&=&Q_i^{(0)}(\tau,\tau_{1})+\epsilon^{2}Q_i^{(1)}(\tau,\tau_{1})+\dots
\\
p_i(\tau,\tau_{1})&=&P_i^{(0)}(\tau,\tau_{1})+\epsilon^{2}P_i^{(1)}(\tau,\tau_{1})+\dots
\end{eqnarray}
in such a form that  the corrections $Q_i^{(1)}$ and $P_i^{(1)}$ must be not secular with respect to $Q_i^{(0)}$ and $P_i^{(0)}$ in $\tau$ time; i.e. $Q_i^{(1)}$ and $P_i^{(1)}$ must be consistently be of order $\epsilon^2$ with respect to $Q_i^{(0)}$ and $P_i^{(0)}$  for all values of $\tau$. 
The derivative symbol ``$\cdot$''  is now replaced by $d/d\tau=\partial_{\tau} +\epsilon^{2}\partial_{\tau_{1}}$.
At leading order, 
it is nearly obvious that the choosing of $Q_{z}^{(0)}$  as 
\begin{equation}
Q_{z}^{(0)}(\tau,\tau_{1}) = q_{z}^{(0)}(\tau_{1}) \equiv 1 + c(\gamma)\tau_{1}\, , 
\end{equation}
removes the linear secularity. 
An explicit computation with the assumed initial conditions yields the following implicit expression
\begin{equation}
Q_\perp^{(0)}
{}_2F_1\left[-\frac{1}{2\gamma},\frac{1}{2}, 1-\frac{1}{2\gamma};\frac{{q_{z}^{(0)}}^{-2\gamma}}{{Q^{(0)}_{\perp}}^{2\gamma}}\right] = \frac{\tau}{\sqrt{\gamma}}+
\frac{(\gamma-1)c(\gamma)}{\sqrt{\gamma}} ,   
\end{equation}
which produces the asymptotic behavior 
\begin{eqnarray}
Q_\perp^{(0)}(\tau, \tau_1) &\sim& \frac{\tau + (\gamma-1) c(\gamma)}{\sqrt{\gamma}}
\\\nonumber
&&\qquad - 
\frac{\gamma^{\gamma - \tfrac{1}{2}} \left(q_{z}^{(0)}(\tau_1)\right)^{-\gamma}  \tau^{1-2\gamma} }{2 - 4 \gamma},  \\ 
P_\perp^{(0)}(\tau, \tau_1) &\sim& \frac{1}{\sqrt{\gamma}} \left[1  - \frac{1}{2} 
\gamma^{\gamma - \tfrac{1}{2}}  \left(q_{z}^{(0)}(\tau_1)\right)^{-\gamma}  \tau^{-2\gamma}  \right], \\ 
P_z^{(0)}(\tau, \tau_1) &\sim&   \frac{c(\gamma)}{2}   
+ \frac{\gamma^{\gamma} \left(q_{z}^{(0)}(\tau_1)\right)^{-1-\gamma}  \tau^{1-2\gamma} }{2 - 4 \gamma},
\end{eqnarray}
for $\tau \to \infty$ at fixed $\tau_{1}$ ($\tau_1 = O(1)$).
Now the equation for $Q_z^{(1)}$ is simply 
\begin{equation}
\frac{\partial Q_z^{(1)}}{\partial \tau} = -c(\gamma) + 2 P_z^{(0)}
\end{equation}
which in the region $\tau \to \infty,\,  \tau_1 = O(1)$ is of order $\tau^{1-2\gamma}$.  
This  leads to an asymptotic  growth rate that is slower than linear.

Let us now consider the cases $\gamma=1$ and $1/2<\gamma<1$ separately. For the boson case $\gamma=1$  the expressions are more transparent and become
\begin{eqnarray}
Q_\perp^{(0)}(\tau, \tau_1) &=& \frac{\sqrt{2 + \tau^2(2+\pi \tau_1)}}{\sqrt{2 + \pi \tau_1}} , \\ 
P_\perp^{(0)}(\tau, \tau_1)&=& \frac{\tau \sqrt{2 + \pi \tau_1}}{\sqrt{2 + \tau^2(2+\pi \tau_1)}} , \\ 
P_z^{(0)}(\tau, \tau_1) &=&  \frac{\sqrt{2}}{(2+\pi \tau_1)^{3/2}} \arctan\left(\tau \sqrt{1 + \tfrac{\pi}{2} \tau_1}\right), \\
Q_{z}^{(1)}(\tau,\tau_{1})&=&
 \frac{\tau}{(1+\tfrac{\pi}{2} \tau_1)^{3/2}}\left[ \arctan\left(\tau \sqrt{1 + \tfrac{\pi}{2} \tau_1}\right) -\frac{\pi}{2}\right]
 \nonumber\\
 &&-\frac{\ln\left[1 + \tau^2(1 + \pi \tau_1/2)\right]}{2 \left(1 + \tfrac{\pi}{2} \tau_1\right)^2}  . 
 \end{eqnarray}
Therefore, we see that $Q_{z}^{(1)}(\tau,\tau_{1})$  
 only grows logarithmically as $\tau\to\infty$ and $\tau_1$ is $O(1)$,  in contrast 
 with the naive perturbative result  of  Eq.~(\ref{eq:bperp_cd}).
 
We can exploit  these results  to improve the perturbative evaluation of  the aspect ratio to higher order in $\epsilon$. Notice that earlier we were able to predict the asymptotic ratio to first order from the zeroth order result by using the Hamiltonian approach; analogously here we can obtain an improved aspect ratio.

Let us now consider the axial momentum, whose asymptotic value can be obtained as 
\begin{equation}
p_{z}^{\infty}=\int_{0}^{\infty}d\tau\dot{p}_z(\tau,\tau_{1}(\tau));
\end{equation}
in order to extract the next-to-leading corrections, it is sufficient to consider its equation of motion
up to order $\epsilon^{2}$
\begin{equation}
\label{eq:dotpz}
\dot{p}_z=\frac{1}{2 {Q^{(0)}_{\perp}}^{2}{q^{(0)}_z}^{2}}
-\epsilon^{2}\left(
\frac{ Q^{(1)}_{\perp}}{{Q^{(0)}_{\perp}}^{3}{q^{(0)}_z}^{2}}
+\frac{Q^{(1)}_{z}}{{Q^{(0)}_{\perp}}^{2}{q^{(0)}_z}^{3}}
\right)+\dots
\end{equation}
Then, the integration of the first term gives
\begin{eqnarray}
&&\int_{0}^{\infty}\frac{d\tau}{2 Q^{(0)}_{\perp}(\tau,\tau_{1}(\tau))^{2}q^{(0)}_z(\tau_{1}(\tau))^{2}}
\nonumber\\
&&\sim
\frac{\pi}{4}+\frac{\pi\epsilon^{2}}{8}\left[1  + 4 \ln\left( \frac{\pi\epsilon^{2}}{2}\right)\right]+\dots\nonumber, \quad \epsilon \to 0 , 
\end{eqnarray}
where the ``resummation''  has been crucial to ensure the convergence of the integral.
For the other two terms, we may neglect the dependence on $\tau_{1}$ (since it introduces corrections of order $\epsilon^2$ in the integral, which compete with higher order terms in the corrected perturbative expansion), 
by using $q^{(0)}_z(0) =1$,  $Q^{(1)}_{z}(\tau ,0)=\tau\arctan\tau - \ln\sqrt{1+\tau^2}-\frac{\pi}{2}\tau$, and $Q^{(0)}_{\perp}(\tau,0) = \sqrt{1+\tau^2}$. 
Therefore $Q^{(1)}_{\perp}(\tau ,0) \equiv b_\perp^{(1)}(\tau)$ is exactly the first non-resummed radial correction satisfying 
\begin{equation}
\label{eq:bperp1}
\ddot{b}_\perp^{(1)}+\frac{3b_\perp^{(1)} }{{b_\perp^{(0)}}^4}= -\frac{b_z^{(1)}}{{b_\perp^{(0)}}^3} 
\end{equation}
where $b_z^{(1)}(\tau)=\tau\arctan\tau - \ln\sqrt{1+\tau^2}$ (see Eq. (\ref{eq:bperp_cd}).
The solution of Eq. (\ref{eq:bperp1}) with zero initial conditions may be written  as
\begin{equation}
b_\perp^{(1)} (\tau)=
-\int_{0}^{\tau}G(\tau,s)\frac{ b_z^{(1)} (s)}{(1+s^2)^{3/2}}ds 
\end{equation}
with $G(\tau,s)$ being the following Green's function 
\begin{equation}
G(\tau,s)=\theta(\tau-s)\frac{\tau-s}{\sqrt{1+\tau^{2}}}\frac{1+s\tau}{\sqrt{1+s^{2}}}\, .
\end{equation}
The corresponding integrations produce 
 \begin{equation}
-\int_{0}^{\infty}\!\!\!\!\!d\tau\left(\frac{ Q^{(1)}_{\perp}(\tau,0)}{{Q^{(0)}_{\perp}}(\tau,0)^{3}}
+\frac{Q^{(1)}_{z}(\tau,0)}{{Q^{(0)}_{\perp}}(\tau,0)^{2}}\right)
=\frac{\pi}{8}(1-\ln2)+\pi\ln2 . 
\end{equation}
and 
\begin{equation}
 \dot{b}_\perp^{(1)}(\infty) = -\frac{\pi^2}{16}, 
 \end{equation}
 that corresponds to the first order correction to  the asymptotic radial momentum.
 \begin{figure}[b!]
\includegraphics[width=0.7\columnwidth]{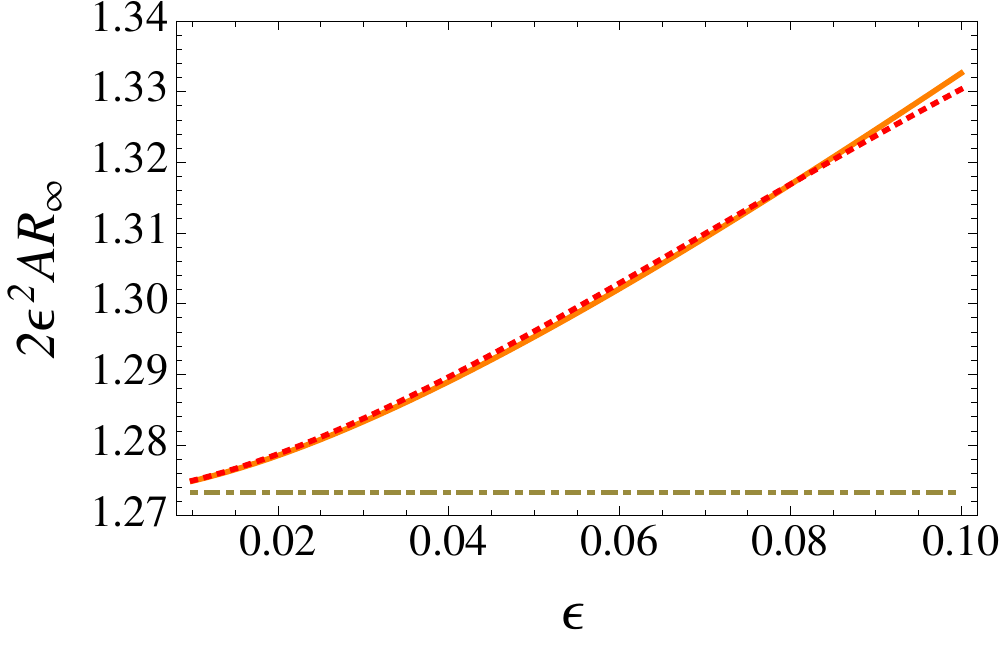}
\caption{(Color online) Asymptotic aspect ratio $2\epsilon^2 AR_{\infty}$ as a function of $\epsilon^{2}$, for $\gamma=1$. The full numerical solution of Eqs. (\ref{eq:bz})-(\ref{eq:bperp}) (orange solid line) is compared with the zeroth-order expression in Eq. (\ref{eq:asympt}) (dot-dashed line) and with those including next-to-leading corrections for $\gamma=1$ (red dotted line).}
\label{fig:asympt1}
\end{figure}

Thus, one obtains the asymptotic values 
\begin{eqnarray}
p_{\perp}^{\infty}&=& 1 -\frac{\pi^2 \epsilon^2}{16}  + \ldots, \\ 
p_{z}^{\infty}&=&
\frac{\pi}{4}+\frac{\pi\epsilon^{2}}{8}\left( 2 + 3 \ln 2 + 4 \ln(\pi\epsilon^{2})\right) + O(\epsilon^{4}\ln\epsilon) . 
\label{eq:pz1}
\end{eqnarray}
These results represent a significative improvement for the asymptotic ratio $AR_{\infty}=(1/2\epsilon^2)(p_\perp/p_z)_{\infty}$ with respect to the zeroth-order result in Eq. (\ref{eq:gamma1-0}), as  shown in Fig. \ref{fig:asympt1} (red dotted line), and in the three panels of Fig. \ref{fig:asympt} (red dots). 

Finally, let us now consider briefly the asymptotics of momenta in  the case $1/2<\gamma<1$.  
Now 
$p_z^\infty$ is given by
\begin{equation}
p_z^\infty \sim \int_{0}^{\infty}d\tau\frac{1}{2 {Q^{(0)}_{\perp}(\tau,0)}^{2\gamma}{q^{(0)}_z(\tau_{1}(\tau))}^{\gamma+1}} + O(\epsilon^2) ,
\end{equation}
where the $O(\epsilon^2)$ terms  come from integrations that are analogous to those in the last  brackets of  Eq.~(\ref{eq:dotpz}). 
It is possible to perform an asymptotic expansion of above  integral and  to obtain the two leading terms. 
The result is 
\begin{widetext}
\begin{eqnarray}
p_{z}^{\infty}&\sim &\frac{c(\gamma)}{2} +\frac{\gamma^{\gamma-1 } c(\gamma)^{2\gamma-1}}{6(1-2\gamma)} 
 \epsilon^{2 (2\gamma-1)} \left[3 \gamma  \, _2F_1 (1+\gamma ,1-2 \gamma ,2-2 \gamma ,-1)+
      (1-2 \gamma )  \,_2F_1 (1+\gamma ,3 \gamma ,1+3 \gamma ,-1)\right]  \nonumber  \\ 
      && + \frac{\sqrt{\pi}\, \gamma (1+\gamma) c(\gamma)   \Gamma\left(2-\tfrac{1}{\gamma} \right)}  
      {4(1-\gamma) \Gamma\left(\tfrac{3}{2}-\tfrac{1}{\gamma} \right)} \epsilon^2   + O\left( \epsilon^2\right) . 
\label{eq:pz2}
\end{eqnarray}
\end{widetext}
Notice that the expansion contains a term proportional to $(\epsilon^2)^{2 \gamma -1}$ 
that dominates over the $\epsilon^2$ term when  $1/2<\gamma<1$. 
However, the latter term cannot be discarded, as it is needed to cancel the singularity in the limit $\gamma\to1$, 
producing an $O(\epsilon^2 \ln \epsilon)$ term.  
Therefore, the improved asymptotic ratio $AR_{\infty}$ can be obtained by combining Eq. (\ref{eq:pz2}) 
with 
\begin{equation}
p_\perp^\infty =  \frac{1}{\sqrt{\gamma}}  - \frac{c(\gamma)^2 \sqrt{\gamma}}{4} \epsilon^2 + \ldots , 
\end{equation}
where the second term is easily obtained from energy conservation. 
Even in this case, it represents a significant correction with respect to the zeroth-order result, as shown in Fig. \ref{fig:asympt}. 

\begin{figure}[h!]
\includegraphics[width=0.7\columnwidth]{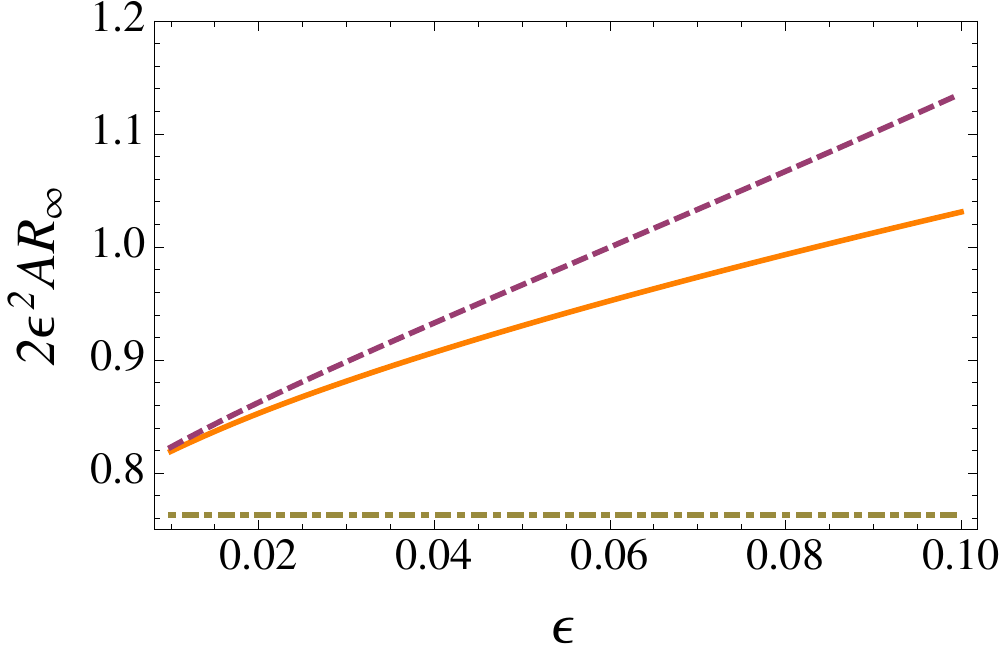}
\caption{(Color online) Asymptotic aspect ratio $2\epsilon^2 AR_{\infty}$ as a function of $\epsilon^{2}$, for $\gamma=2/3$. The full numerical solution of Eqs. (\ref{eq:bz})-(\ref{eq:bperp}) (orange solid line) is compared with the zeroth-order expression in Eq. (\ref{eq:asympt}) (dot-dashed line) and with that including next-to-leading corrections (dashed line).}
\label{fig:asympt23}
\end{figure}

In particular, the asymptotic expressions for the fermionic case, $\gamma=2/3$, are 
\begin{eqnarray}
p_{z}^{\infty} &=& 1.6057 - 2.5333\, \epsilon^{2/3} + \ldots , \\ 
p_{\perp}^{\infty} &=& 1.2248 - 2.1051\, \epsilon^2 + \ldots
\end{eqnarray}
and the corresponding aspect ratio is shown in in Fig. \ref{fig:asympt23}, as a function of $\epsilon^{2}$. In this case, the improvement provided by the  multiple-scale approach  over the naive zeroth-order result is quite significant in the small $\epsilon$ regime.

\section{Conclusions}
The asymptotic aspect ratio predicted from the naive expansion is surprisingly adequate, when considering that the naive expansion is invalid for the large times for which the aspect ratio is desired. We have explained this success as being derived from the Hamiltonian character of the equations of motion, so that in fact it is not a perturbative result. We have also produced resummed corrections to next order, in which the non-analycity becomes apparent. The techniques applied here have a long history in applied mathematics;  most of the published examples of the multiple scales method, however, concern bounded periodic motion. Here we make an novel application thereof to asymptotically linear solutions.  
We have provided perturbative expansions in the anisotropy parameter which are valid over the whole expansion time, and not only for small times. This could be used to extract trap information out of longer time-of-flight experiments.
The techniques presented here have wider applicability, and could be applied to, for instance, expansions of multiple species.
\begin{acknowledgements}

I.L.E and M.A.V. acknowledge funding by the Basque Government (Grant No. IT559-10) and the Spanish Ministry of Science and Technology (Grant No. FPA2009-10612 and Consolider-Ingenio 2010 Programme CPAN CSD2007-00042).\end{acknowledgements}


\begin{thebibliography}{}
\bibitem{nota}
To get the proper aspect ratio one has to multiply $AR_{\infty}$ by $R_\perp(0)/R_z(0)$.
\end{thebibliography}
\end{document}